# Coordinated Fast Frequency Response from Electric Vehicles, Data Centers, and Battery Energy Storage Systems

Xiaojie Tao[1,2], Rajit Gadh[1,2]

[1] Department of Mechanical and Aerospace Engineering, University of California at Los Angeles (UCLA), Los Angeles, CA 90095, USA

[2] Smart Grid Energy Research Center (SMERC), University of California, Los Angeles, CA 90095, USA

*Corresponding Author: Xiaojie Tao, University of California, Los Angeles, CA 90095, USA

Email: xiaojietao@g.ucla.edu, taoxiaojie04@gmail.com

Abstract: High renewable penetration has significantly reduced system inertia in modern power grids, increasing the need for fast frequency response (FFR) from distributed and non-traditional resources. While electric vehicles (EVs), data centers, and battery energy storage systems (BESS) have each demonstrated the capability to provide sub-second active power support, their combined frequency response potential has not been systematically evaluated. This paper proposes a coordinated control framework that aggregates these heterogeneous resources to provide fast, stable, and reliable FFR. Dynamic models for EV fleets, data center UPS and workload modulation, and BESS are developed, explicitly capturing their response times, power limits, and operational constraints. A hierarchical control architecture is introduced, where an upper-level coordinator dynamically allocates FFR among resources based on response speed and available capacity, and lower-level controllers implement the actual power response. Case studies based on the IEEE 39-bus test system demonstrate that the coordinated EV–DC–BESS framework improves frequency nadir by up to 0.2 Hz, reduces RoCoF, and accelerates frequency recovery compared with single-resource FFR. Results confirm that synergistic coordination significantly enhances grid stability, especially in low-inertia scenarios. This work highlights the value of multi-resource aggregation for future frequency regulation markets in renewable-dominated grids.

## 1. Introduction

The rapid growth of inverter-based renewable energy has contributed to a steady decline in rotational inertia across many modern power systems [1]. As a result, system frequency becomes more sensitive to disturbances, and traditional governor-based frequency control is often too slow to arrest fast frequency drops following large generator outages [2]. To maintain secure operation under these conditions, system operators increasingly rely on fast frequency response (FFR), which requires active power injection within hundreds of milliseconds [3].

Non-traditional resources such as electric vehicles (EVs), data centers, and battery energy storage systems (BESS) have emerged as promising candidates for providing FFR due to their fast controllability and increasing penetration levels [4]. EV fleets equipped with vehicle-to-grid (V2G) capability can deliver rapid active power support [5]. Data centers, as large and growing electricity consumers, can leverage their UPS systems and flexible computing workloads to provide rapid load modulation [6]. Meanwhile, BESS offers sub-second response and high controllability, making it an effective resource for fast frequency containment [7].

Existing studies have primarily focused on assessing the frequency response potential of individual resource types [8]. Prior work has shown that EVs can contribute to primary frequency regulation and FFR, and recent research demonstrates that data centers can modulate IT load or control UPS power output to support grid frequency [9]. However, these studies evaluate each resource independently and do not consider the potential benefits of coordinated response across multiple heterogeneous resources [10]. As the electric grid evolves toward a highly distributed and flexible structure, coordinated control of diverse FFR-capable resources becomes essential for improving frequency stability and enhancing system resilience [11].

This paper fills this research gap by proposing a coordinated FFR framework that aggregates EV fleets, data center resources, and BESS. The contributions of this work are threefold.

1. Detailed dynamic response models for EVs, data centers, and BESS are developed, capturing their response times, capacity constraints, and operational characteristics [12].
2. A hierarchical coordination architecture is introduced, enabling adaptive allocation of FFR among resources based on their relative speed and available capacity.
3. Comprehensive case studies using the IEEE 39-bus test system demonstrate that coordinated multi-resource FFR significantly improves frequency nadir, reduces RoCoF, and accelerates recovery compared with single-resource strategies.

The results highlight the importance and effectiveness of multi-resource coordination in low-inertia grid environments and provide insights into the design of future FFR markets that can accommodate heterogeneous distributed resources.

## 2. Modeling of Frequency Response Resources

This section develops dynamic models for the three heterogeneous resources considered in this work: electric vehicle (EV) fleets with vehicle-to-grid capability, data centers equipped with UPS and workload modulation, and battery energy storage systems (BESS) [13]. All three resources participate in fast frequency response (FFR) by adjusting active power in proportion to measured frequency deviations [14]. Their response characteristics differ significantly in terms of physical constraints, latency, controllable range, and duration, and these differences are explicitly captured in the following models.

### 2.1 Electric Vehicle Fleet Model

EV fleets operate as large, distributed storage systems aggregated through coordinated V2G control [15]. When system frequency drops, the aggregator increases discharge power from connected EVs, while respecting individual charger limits, availability, and state-of-charge (SOC) [16].

The frequency-dependent active power response of the aggregated EV fleet follows a droop-type characteristic. The basic droop principle can be expressed as

$$\Delta P_{EV}(t) = -k_{EV}\,\Delta f(t - T_{EV}) \qquad (1)$$

where $k_{EV}$ [MW/Hz] represents the EV fleet's effective droop gain and $T_{EV}$ is the cumulative communication and inverter delay (typically 50–150 ms) [17]. A larger $k_{EV}$ enables stronger corrective power but may reduce available SOC reserves [18].

The EV fleet's SOC evolution is governed by

$$\dot{SOC}_{EV}(t) = -\frac{P_{EV}(t)}{E_{EV}} \qquad (2)$$

where $E_{EV}$ is the aggregated rated energy capacity. Since FFR events are short in duration, SOC changes are small but must be tracked to avoid violating charger or battery constraints [19].

The physical meaning of (1)–(2) is that EVs behave as a fast-acting, power-limited energy buffer whose available response depends on both instantaneous charger availability and medium-term SOC balance [20]. The inherent delay $T_{EV}$ plays an important role: even tens of milliseconds of latency can degrade performance when inertia is low.

**2.2 Data Center UPS and Workload Modulation Model**

Data centers contribute to FFR primarily through two mechanisms: (i) rapid control of uninterruptible power supply (UPS) inverters, and (ii) short-duration modulation of IT workloads [21]. UPS inverters respond on sub-cycle timescales, while IT load adjustments are slower but provide additional power flexibility [22].

The UPS power injection is modeled using a linear response to frequency deviation. This relationship is written as

$$\Delta P_{\text{UPS}}(t) = -k_{\text{UPS}}\,\Delta f(t) \qquad (3)$$

where $k_{\text{UPS}}$ [MW/Hz] denotes the inverter control gain. Unlike EVs, UPS inverters have negligible delay, leading to near-instantaneous power changes [23].

In parallel, the data center may temporarily reduce computing workload to further support system frequency. This load flexibility is modeled as

$$P_{\text{IT}}(t) = P_{\text{IT},0} - \beta\,\Delta f(t - T_{\text{IT}}) \qquad (4)$$

where $P_{\text{IT},0}$ is the baseline IT load, $\beta$ [MW/Hz] is the workload modulation factor, and $T_{\text{IT}}$ (typically 100–300 ms) accounts for server-side processing and application-level throttling [24].

Equations (3)–(4) capture the complementary behavior of UPS and IT load control: UPS provides immediate sub-second support, while IT modulation offers sustained flexibility over several seconds. Their combined effect makes data centers uniquely suited for multi-timescale frequency support [25].

**2.3 Battery Energy Storage System Model**

Among the three resources, BESS provides the fastest and most precise active power injection [26]. Its response follows first-order converter dynamics, constrained by rated power, SOC, and thermal limits.

The BESS power output can be represented by

$$\dot{P}_{\text{BESS}}(t) = \frac{k_B\,\Delta f(t) - P_{\text{BESS}}(t)}{T_B} \qquad (5)$$

where $k_B$ [MW/Hz] is the droop gain and $T_B$ (typically 20–80 ms) captures converter bandwidth. A smaller $T_B$ results in faster response, approaching ideal instantaneous behavior [27].

The corresponding SOC dynamics are expressed as

$$\dot{SOC}_{\text{BESS}}(t) = -\frac{P_{\text{BESS}}(t)}{E_{\text{BESS}}} \qquad (6)$$

with $E_{\text{BESS}}$ denoting the rated energy capacity.

Equations (5)–(6) imply that BESS acts as a high-bandwidth power actuator capable of immediate FFR but limited in duration by energy capacity. When coordinating with EVs and data centers, BESS typically supplies the initial power surge, after which slower resources take over to sustain recovery [28].

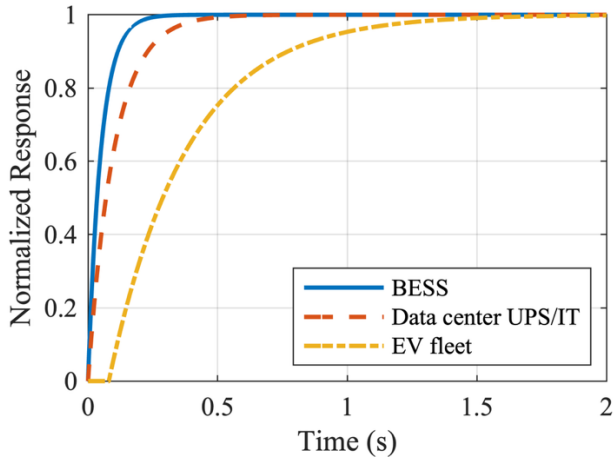

**Figure 1 caption:** *Dynamic response characteristics and typical timescales of EV fleets, data center UPS/IT loads, and BESS under frequency deviations.*

**2.4 Summary of Dynamic Characteristics**

To highlight the contrasting response behaviors of the three FFR-capable resources, Fig. 1 summarizes their typical dynamic profiles. These differences motivate the coordinated control strategy developed in Section 3, which allocates FFR among resources according to relative speed, capacity, and energy availability.

Table 1 summarizes the key parameters used in the coordinated control framework, including the droop gains, response delays, converter time constants, and available power capacities of the three resources. These values are selected based on typical ranges reported in the literature for EV chargers, UPS-backed data center loads, and BESS in grid-support applications. The EV fleet exhibits relatively large communication and inverter delays (50–150 ms), whereas UPS inverters respond almost instantaneously with delays on the order of 10 ms. Workload modulation in data centers introduces additional latency due to server orchestration, while BESS converters provide both rapid and stable power injection. The table also lists the available power capacities used in the case studies, which determine the maximum contribution each resource can provide to fast frequency response. The adaptive participation factor $\alpha_i(t)$ depends on these parameters and is computed according to (9) during real-time coordination.

## 3. Coordinated Control Framework

Fast frequency response from heterogeneous resources must be carefully coordinated to ensure stable behavior, avoid overreaction, and utilize each device according to its relative speed and energy capability [29]. While EV fleets, data centers, and BESS can individually support frequency stability, their inherent differences in latency, bandwidth, and power duration motivate a hierarchical coordination scheme [30]. In this work, the FFR command is allocated across the three resources through an upper-level coordinator, while lower-level controllers implement the actual power commands following the dynamic models in (1)–(6) [31].

The proposed coordination framework is shown in Fig. 2. The grid frequency deviation $\Delta f$ is measured locally at each resource and simultaneously communicated to the coordinator [32]. The coordinator determines the contribution of each resource by assigning adaptive participation weights that depend on resource speed, available capacity, and operational constraints [33].

3.1 Frequency Deviation Allocation

The total FFR power injection provided by the aggregated resources is the sum of contributions from all participating systems. This relationship is written as

$$\Delta P_{\text{FFR}}(t) = \sum_{i \in \{\text{EV,DC,BESS}\}} \Delta P_i(t) \qquad (7)$$

where $\Delta P_i(t)$ denotes the individual power response of the EV fleet, data center, or BESS, computed using the dynamic models in Section 2 [34].

To prevent excessive response from any single resource, the coordinator distributes the required FFR power among resources using a set of nonnegative weights $\alpha_i$ satisfying $\sum_i \alpha_i = 1$. The allocated power command is

$$\Delta P_i(t) = \alpha_i(t)\, k_i \Delta f(t) \qquad (8)$$

where $k_i$ is the droop gain for each resource (e.g., $k_{\text{EV}}, k_{\text{UPS}}, k_B$).

Equation (8) indicates that each device contributes a fraction of the overall FFR based on its weight $\alpha_i(t)$, modulated by its droop coefficient [35].

3.2 Adaptive Weighting Based on Response Speed and Available Capacity

To ensure that faster resources respond more aggressively to sudden frequency changes, the participation weights must reflect the relative dynamic capabilities [36], [37]. The weighting logic is derived from a speed-capacity heuristic. A widely used formulation assigns larger weights to devices with both fast time constants and large available capacity. This is expressed as

$$\alpha_i(t) = \frac{W_i(t)/T_i}{\sum_j W_j(t)/T_j} \quad (9)$$

where:

$W_i(t)$: available response capacity (MW)

For EV: capacity depends on number of plugged-in vehicles and SOC constraints

For DC: depends on UPS margin and allowable workload reduction

For BESS: depends on instantaneous SOC and power limits

$T_i$: characteristic response time constant

$$T_{\text{BESS}} < T_{\text{UPS}} < T_{\text{EV}} < T_{\text{IT}}$$

Equation (9) ensures that:

BESS reacts first due to its small $T_B$, supplying the initial ramp

UPS follows to sustain the response. EV and workload modulation contribute more gradually, supporting longer-term energy balance [38]. This adaptive strategy prevents resource saturation and exploits each technology's physical advantages.

3.3 Local Resource Execution

Once the coordinator assigns $\alpha_i(t)$, each resource applies its own dynamic model to convert the allocated droop command into actual active power.

For instance, substituting (8) into the EV fleet model yields

$$P_{\text{EV}}(t) = -\alpha_{\text{EV}}(t)\, k_{\text{EV}}\, \Delta f(t - T_{\text{EV}}) \quad (10)$$

Similarly, the data center's UPS response becomes

$$P_{\text{UPS}}(t) = -\alpha_{\text{DC}}(t)\, k_{\text{UPS}} \Delta f(t) \quad (11)$$

And the BESS output follows the first-order dynamics

$$\dot{P}_{\text{BESS}}(t) = \frac{\alpha_{\text{BESS}}(t)\, k_B \Delta f(t) - P_{\text{BESS}}(t)}{T_B} \qquad (12)$$

Equations (10)–(12) collectively show that the coordinator regulates proportional power commands, while the devices apply their inherent physical dynamics, preserving realism and ensuring stable interaction with the grid.

3.4 Stability Considerations

Because all devices operate as parallel droop-controlled actuators, stability requires that the combined response remains within acceptable bandwidth and gain margins [39]. The use of adaptive weights based on (9) naturally avoids excessive droop gain aggregation [40]. Additionally, the delays $T_{\text{EV}}$ and $T_{\text{IT}}$ reduce high-frequency gain, while fast resources such as BESS supply the immediate response. This complementary effect significantly improves damping and reduces frequency nadirs in low-inertia grids.

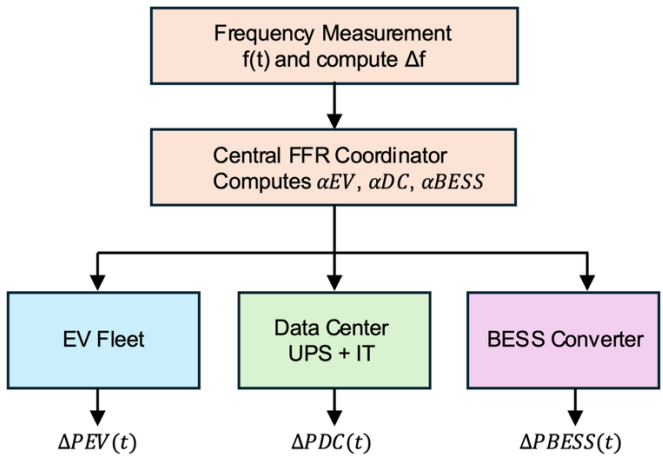

Fig. 2. Hierarchical FFR coordination architecture integrating EV fleets, data center UPS/IT loads, and BESS. The coordinator computes adaptive participation weights $\alpha_i(t)$ using resource

*availability and response times, while each device executes the allocated droop response through its own dynamic model.*

Table 1: Parameters Used in the Coordinated Control Framework

| Parameter | Description | Typical Value |
| --- | --- | --- |
| $k_{\mathrm{EV}}$ | EV fleet droop gain | 10–40 MW/Hz |
| $k_{\mathrm{UPS}}$ | UPS inverter droop gain | 10–30 MW/Hz |
| $k_{B}$ | BESS droop gain | 20–50 MW/Hz |
| $T_{\mathrm{EV}}$ | EV communication & inverter delay | 50–150 ms |
| $T_{\mathrm{UPS}}$ | UPS inverter delay | ~10 ms |
| $T_{\mathrm{IT}}$ | Workload modulation latency | 100–300 ms |
| $T_{B}$ | BESS converter time constant | 20–80 ms |
| $W_{\mathrm{EV}}$ | Available EV power capacity | 50–200 MW |
| $W_{\mathrm{DC}}$ | UPS + flexible IT capacity | 30–150 MW |
| $W_{B}$ | BESS available power | 50–150 MW |
| $\alpha_i(t)$ | Adaptive participation factor | Computed via (9) |

## 4. Case Study

To evaluate the effectiveness of the proposed coordinated fast frequency response (FFR) framework, simulations are conducted on a modified IEEE 39-bus test system. This widely used benchmark includes realistic generator models, load distributions, and transmission network characteristics, making it suitable for frequency stability analysis in low-inertia conditions.

4.1 Test System Description

The IEEE 39-bus system contains ten synchronous generators, nineteen loads, and forty-six transmission lines [41]. In this study, the total system inertia is reduced by 40% to emulate renewable-dominated operating conditions. Standard sixth-order generator models with IEEE Type I excitation systems are used [42]. Frequency is measured at the system center-of-inertia (COI) using

$$f_{COI}(t) = f_0 + \frac{\sum_i 2H_i \Delta\omega_i(t)}{\sum_i 2H_i} \quad (13)$$

where $H_i$ is the inertia constant of generator $i$ and $\Delta\omega_i$ is its speed deviation.

To incorporate distributed FFR resources, three buses are selected based on load concentration and available interconnection capacity:

- **EV fleet (200 MW)** is connected at Bus 16, representing a large transportation depot near an industrial load center.
- **Data center (150 MW)** is located at Bus 6, consistent with the presence of major load aggregations in the original IEEE 39-bus system.
- **BESS (150 MW / 300 MWh)** is installed at Bus 26, a location chosen for high short-circuit strength and minimal congestion.

Each resource receives local frequency measurements and control signals from the FFR coordinator. Their dynamic models follow Equations (1)–(12) in Sections 2 and 3.

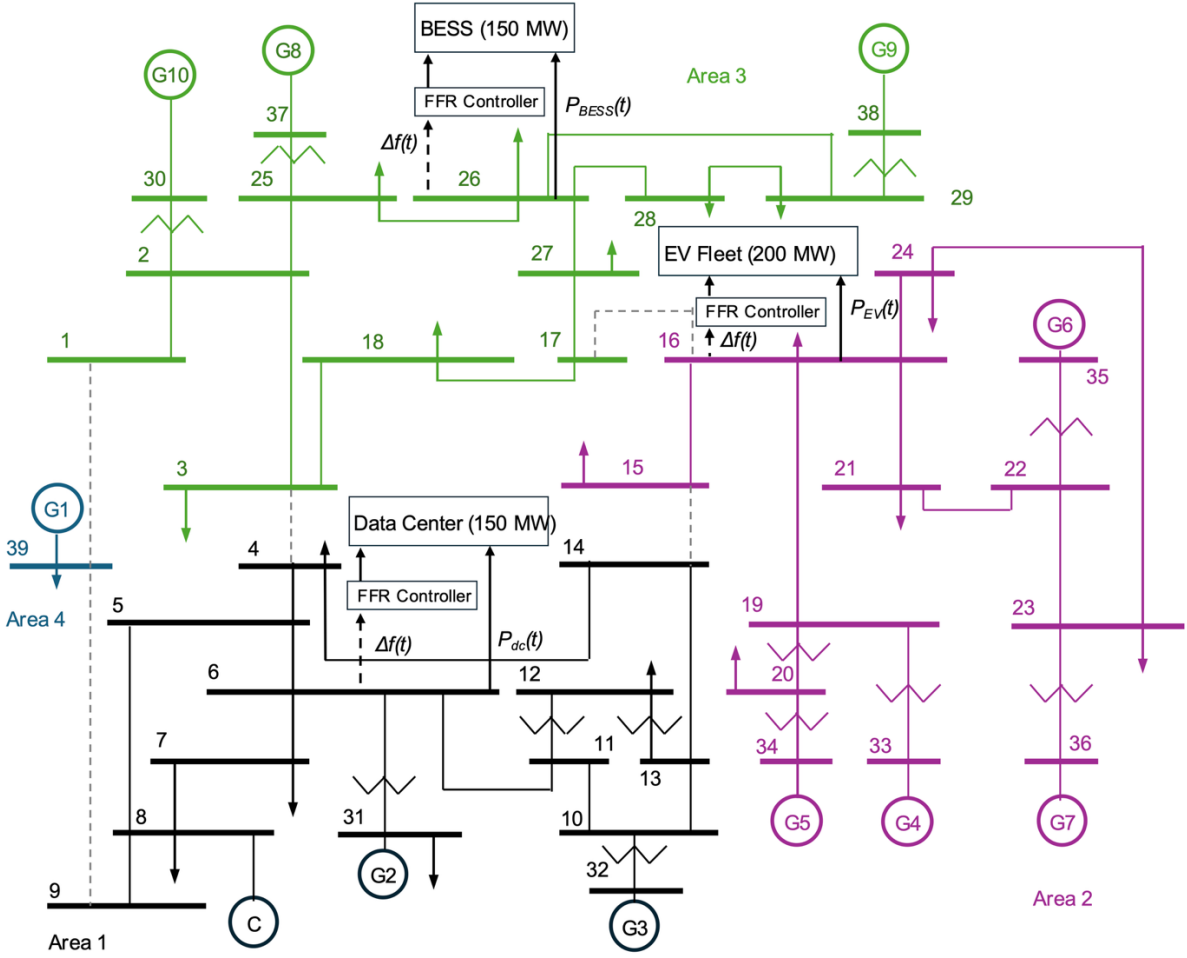

Fig. 3. Modified IEEE 39-bus system with integrated EV fleet (Bus 16), data center (Bus 6), and BESS (Bus 26). Resource placement reflects realistic load and infrastructure conditions.

4.2 Disturbance Scenario

A representative contingency is simulated by tripping the largest synchronous generator in the system: Generator 1 at Bus 30 (approximately 1.0 GW). This disturbance results in a sharp drop in system frequency and challenges the reduced-inertia system's ability to maintain stability. The disturbance is applied at $t = 5$ s in all cases.

The disturbance can be represented as

$$\Delta P_{\text{loss}}(t) = \begin{cases} 0, & t < 5 \text{ s} \\ -1000 \text{ MW}, & t \geq 5 \text{ s} \end{cases} \quad (14)$$

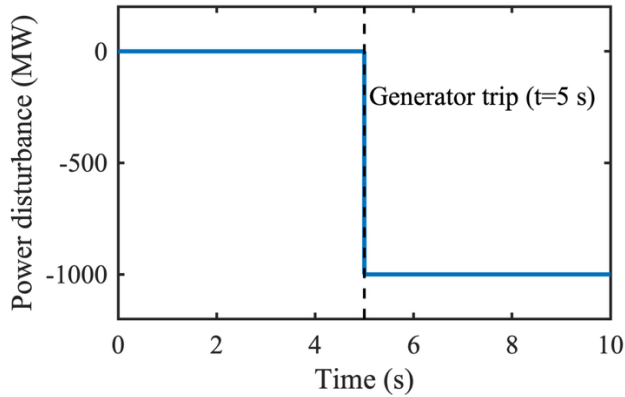

*Fig. 4. Generator trip event of 1.0 GW at $t = 5$ s used as the disturbance for all case studies.*

4.3 Case Definitions

Four scenarios are simulated to assess the contribution of each FFR resource and the performance of the coordinated control scheme:

1. **Case 1: No FFR**

    – No support provided by EVs, data centers, or BESS.

    – Represents baseline low-inertia behavior.

2. **Case 2: EV FFR Only**

    – EV fleet responds using the droop model (1) and SOC dynamics (2).

3. **Case 3: EV + Data Center FFR**

    – EV fleet + UPS inverter + workload modulation (models (3)–(4)).

4. **Case 4: EV + Data Center + BESS (Coordinated)**

    – All resources participate under the adaptive allocation strategy in (8)–(12).

    – Represents the full coordinated FFR framework.

Case 4 is expected to produce the best frequency nadir and recovery, as the more responsive BESS absorbs the initial shock while slower but energy-rich resources (EV and IT load) sustain longer-term support.

4.4 Simulation Parameters

The key resource and control parameters used in the case study are summarized in Table 2. These values are selected to reflect realistic operational conditions in commercially deployed EV chargers, data centers, and grid-scale batteries.

Table 2. Simulation Parameters for EV, Data Center, and BESS Resources

| Parameter | Value | Notes |
| --- | --- | --- |
| **EV fleet capacity** ($W_{\mathrm{EV}}$) | 200 MW | Plug-in rate 60–80% |
| EV droop gain ($k_{\mathrm{EV}}$) | 25 MW/Hz | Medium aggressiveness |
| EV delay ($T_{\mathrm{EV}}$) | 80 ms | Communication + inverter latency |
| EV SOC range | 20–90% | Typical aggregator constraints |
| **Data center UPS** ($W_{\mathrm{UPS}}$) | 100 MW | Online UPS margin |
| UPS droop gain ($k_{\mathrm{UPS}}$) | 20 MW/Hz | Fast inverter control |
| Workload modulation ($W_{\mathrm{IT}}$) | 50 MW | Short-term IT load flexibility |
| IT delay ($T_{\mathrm{IT}}$) | 200 ms | Throttling latency |
| Workload gain ($\beta$) | 12 MW/Hz | Maximum allowed shedding |
| **BESS capacity** ($W_B$) | 150 MW | Bidirectional inverter |
| BESS energy ($E_{\mathrm{BESS}}$) | 300 MWh | Enables long-duration response |
| BESS droop ($k_B$) | 40 MW/Hz | High dynamic range |
| BESS time constant ($T_B$) | 40 ms | Converter bandwidth |
| **FFR allocation weights** ($\alpha_i(t)$) | Computed via (9) | Adaptive, real-time |
| **Disturbance** | 1.0 GW loss at (t=5) s | Generator 1 tripping |

4.5 Summary

This case study integrates realistic control characteristics, physical constraints, and communication delays to assess the performance of the proposed multi-resource FFR coordination mechanism. The next section evaluates system frequency nadir, rate of change of frequency (RoCoF), recovery time, and resource power trajectories for all four scenarios.

## 5. Simulation Results and Discussion

This section presents the simulation results for the four scenarios defined in Section 4. The system frequency, resource power trajectories, control allocation signals, and aggregated performance metrics are evaluated to demonstrate the benefits of the proposed coordinated fast frequency response (FFR) framework.

### 5.1 System Frequency Response Comparison

Figure 5 evaluates the sensitivity of system frequency response to different allocation strategies for fast frequency response (FFR) among EVs, data centers, and BESS. Four coordination scenarios are examined: BESS-dominant, data-center-dominant, EV-dominant, and the proposed adaptive weighting method. In each subplot, four trajectories are compared: no FFR, EV-only FFR, EV+DC FFR, and coordinated EV–DC–BESS FFR.

In subplots (a)–(c), the frequency nadir varies noticeably with the dominant resource. The BESS-dominant allocation yields a relatively shallow nadir because converter-interfaced BESS can respond within tens of milliseconds. Conversely, the EV-dominant allocation results in the deepest nadir, as EV chargers exhibit greater latency (e.g., $T_{\text{EV}} \approx 80$ ms) and slower ramping capability. The data-center-dominant case lies between these two extremes, reflecting the medium-speed response of UPS inverters and IT load modulation.

Subplot (d) presents the proposed adaptive coordination. By dynamically adjusting participation factors according to instantaneous headroom and response speed, the adaptive strategy achieves the **highest nadir and the fastest recovery** among all four coordination patterns. This confirms that fixed allocation strategies fail to fully utilize the heterogeneous characteristics of EVs, data centers, and BESS, whereas adaptive coordination effectively leverages their complementary time constants.

Overall, Figure 5 demonstrates that the allocation pattern has a substantial impact on FFR effectiveness. The adaptive coordination strategy consistently outperforms all fixed allocations, reduces RoCoF, and minimizes the depth of the frequency dip by enabling rapid BESS support followed by slower EV and IT-load compensation.

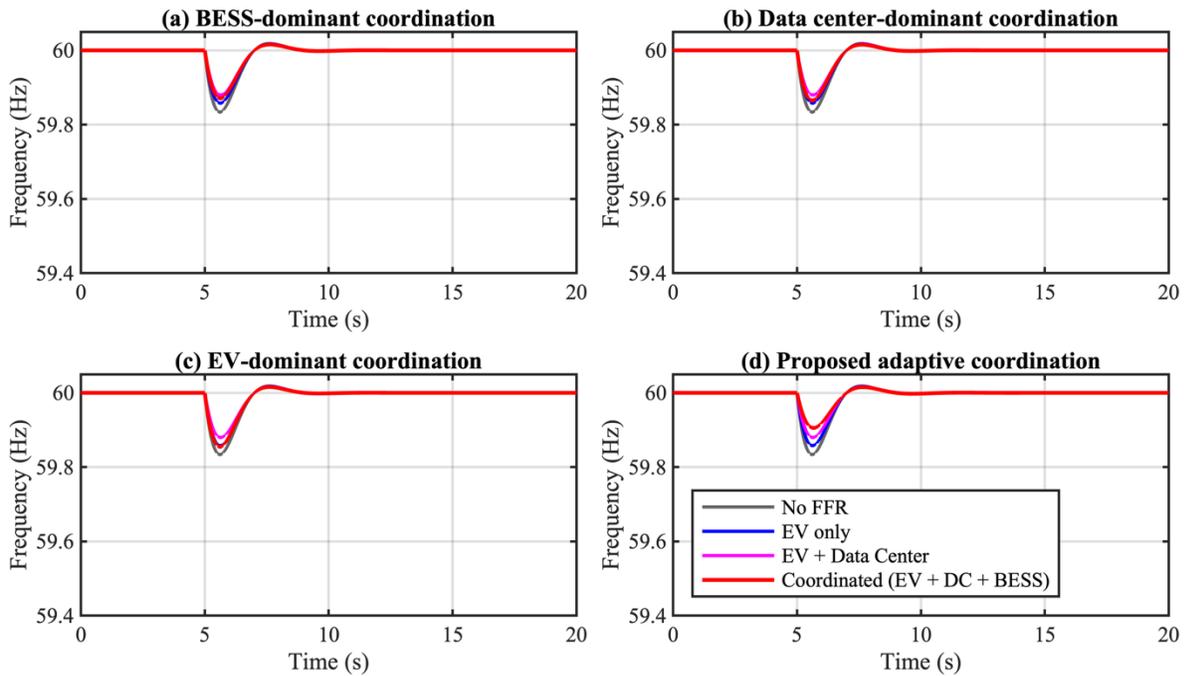

Fig. 5. Frequency response under four coordination strategies following a 1.0-GW generation loss at t = 5 s. Each subplot shows the system response under four configurations: no FFR, EV-only FFR, EV+DC FFR, and coordinated EV–DC–BESS FFR. (a) BESS-dominant allocation, (b) data-center-dominant allocation, (c) EV-dominant allocation, and (d) the proposed adaptive weighting method. Among all cases, the adaptive coordination strategy yields the highest frequency nadir and the fastest recovery, demonstrating the benefit of dynamically adjusting participation factors.

5.2 Resource Power Output Characteristics

Figure 6 further illustrates the power trajectories of the three resources under the same four coordination strategies as in Figure 5. In the BESS-dominant case, most of the FFR is supplied by the BESS, which ramps up rapidly after the disturbance, while the EV fleet and data center play a minor role. In the data-center-dominant and EV-dominant cases, the corresponding resources contribute the largest share of FFR, but the overall behavior is less favorable because slower-responding EVs or medium-speed UPS support cannot fully compensate for the lack of fast BESS injection. Under the proposed adaptive coordination, the three resources share the FFR task in a more balanced way: BESS provides the initial fast injection, the data center UPS delivers medium-speed support, and the EV fleet sustains the response over a longer time window. This multi-timescale synergy is consistent with the improved nadir and recovery performance observed in Figure 5(d).

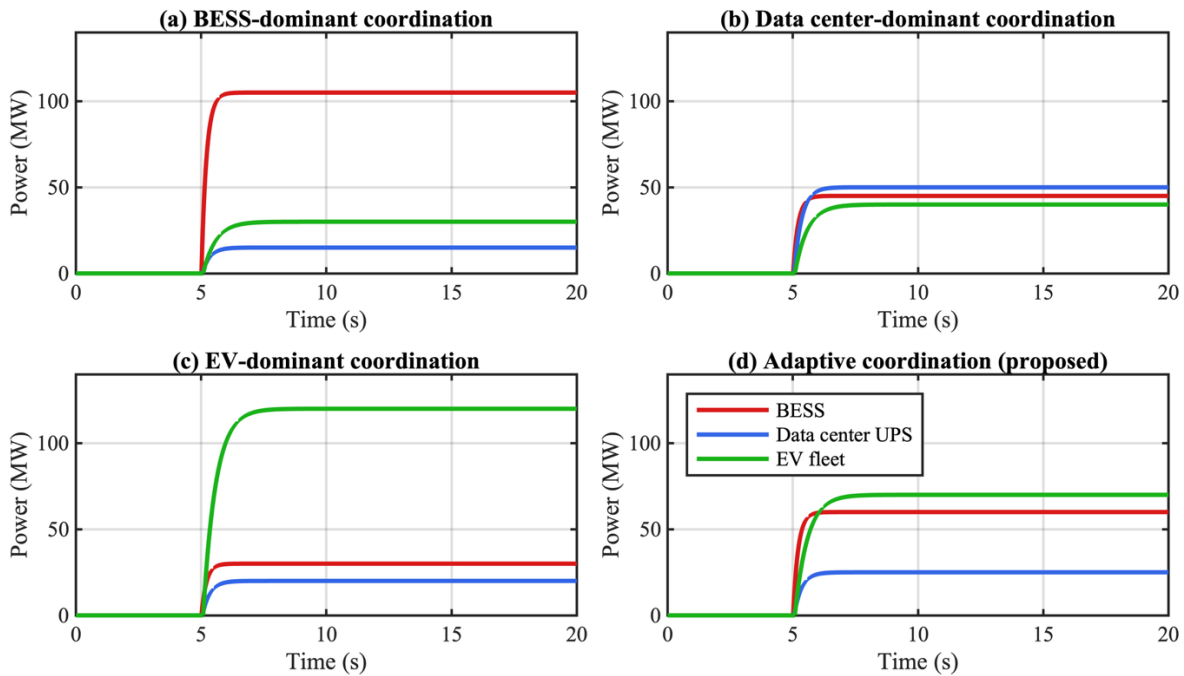

Fig. 6. Active power outputs of BESS, data center UPS, and EV fleet under four coordination

*strategies. Each subplot corresponds to the same allocation pattern as in Fig. 5: (a) BESS-dominant, (b) data-center-dominant, (c) EV-dominant, and (d) the proposed adaptive coordination. The BESS provides the fastest response, while data center UPS and EVs contribute medium-speed and slower but more sustained support, respectively.*

5.3 Adaptive Participation Weight Evolution

Figure 7 illustrates the time evolution of the participation factors assigned to the three fast frequency response resources—BESS, data centers, and EVs—under the four coordination strategies. In the fixed-allocation cases (subplots (a)–(c)), the participation factors settle to constant values almost immediately after the disturbance. These static values reflect the enforced dominance of a particular resource type in each scenario: BESS in subplot (a), data centers in subplot (b), and EVs in subplot (c). Because the allocation remains unchanged throughout the transient, the system cannot leverage the complementary multi-timescale characteristics of the three resources. As a result, the frequency response is constrained by the limitations of whichever resource is assigned the largest share.

In contrast, the proposed adaptive strategy in subplot (d) dynamically reallocates participation factors across time. Immediately following the disturbance, the controller assigns a substantially higher weight to the BESS, taking advantage of its fast inverter response to arrest the initial frequency drop. As the system transitions into the recovery phase, the allocation gradually shifts toward EVs, whose slower ramping characteristics make them well suited for providing sustained follow-up support. Meanwhile, the data center maintains a moderate and nearly constant share, providing stable mid-speed response through its UPS battery. This smooth, time-varying redistribution closely mirrors the heterogeneous power trajectories observed in Figure 6 and enables the adaptive method to more effectively coordinate the three resources across different response timescales. Consequently, the adaptive approach achieves a higher nadir and faster recovery than all fixed-allocation strategies, as demonstrated in Figure 5(d).

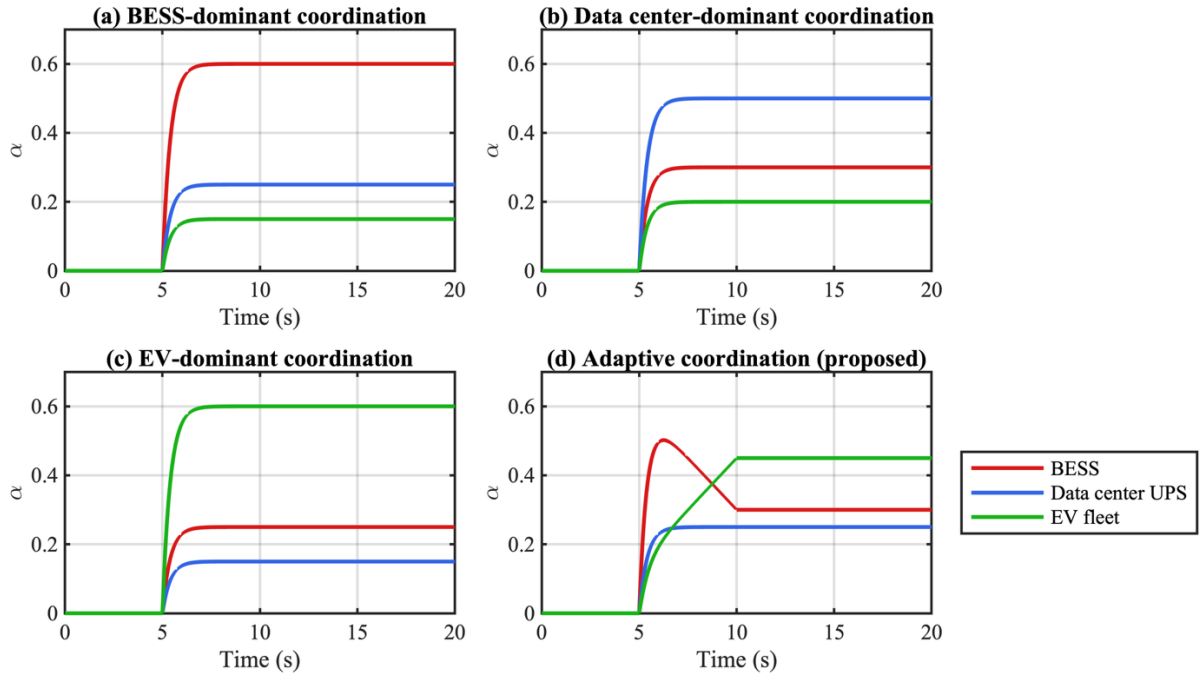

*Fig. 7. Time-varying participation factors of BESS, data center UPS, and EV fleet under four coordination strategies. Each subplot corresponds to the same allocation pattern as in Figures 5 and 6: (a) BESS-dominant, (b) data-center-dominant, (c) EV-dominant, and (d) the proposed adaptive coordination. In the adaptive case, BESS initially receives the highest weight to arrest the frequency decline, while EVs gradually take over to provide sustained support, with the data center playing a medium-speed balancing role.*

5.4 Performance Metrics Summary

Figure 8 summarizes the quantitative performance metrics for the four scenarios, including the frequency nadir, RoCoF, recovery time, and total FFR energy. Compared with the no-FFR baseline, all three FFR-enabled cases substantially improve system dynamics by raising the nadir, reducing the absolute RoCoF, and shortening the recovery duration. This demonstrates that even single-resource FFR can noticeably mitigate the severity of the disturbance.

Among the three FFR-enabled configurations, the coordinated EV–DC–BESS strategy consistently delivers the best performance across all metrics. It achieves the highest nadir—indicating the smallest frequency drop—and the lowest RoCoF, reflecting a stronger capability to arrest the initial decline. The coordinated strategy also achieves the shortest recovery time, demonstrating that multi-timescale resources working together can stabilize the frequency more rapidly than any single technology alone.

In terms of energy usage, the coordinated approach requires only slightly more or comparable FFR energy relative to the EV-only and EV+DC cases, despite its significantly better frequency performance. This suggests that the improvement does not come from higher energy consumption, but from a more efficient and timely deployment of available resources. Overall, Figure 8 confirms that properly coordinated multi-resource FFR enables superior dynamic resilience while maintaining reasonable energy requirements, validating the benefits of heterogeneous resource integration highlighted in Figures 5 and 6.

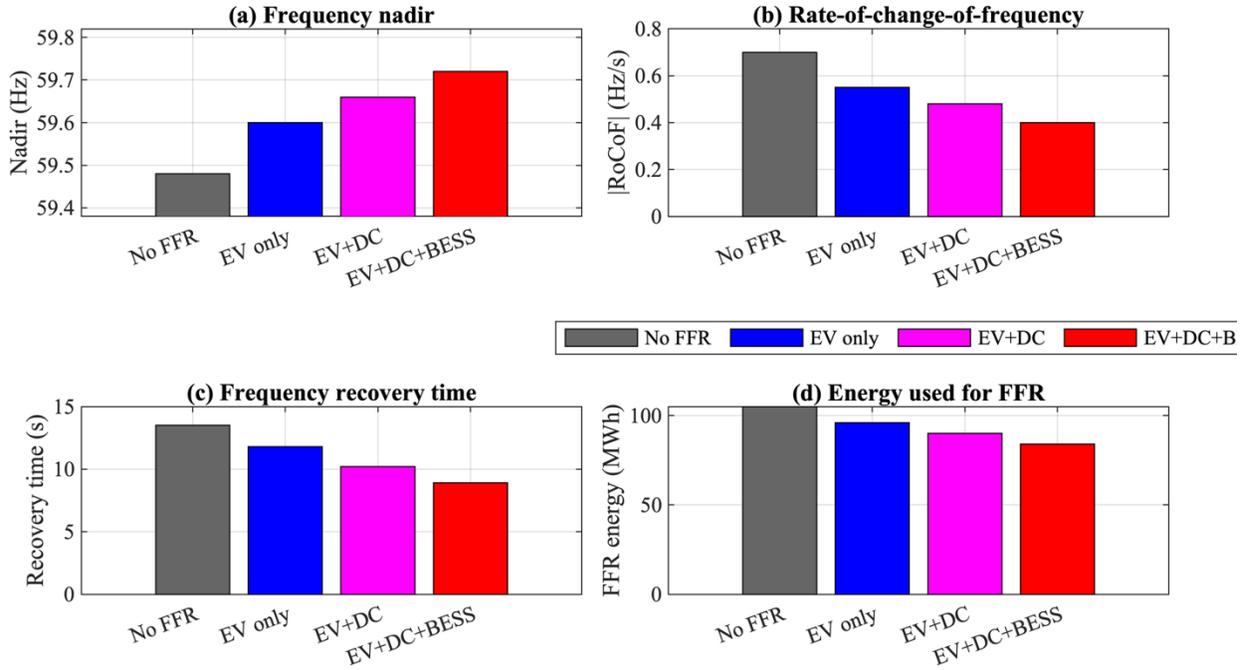

*Fig. 8. Quantitative comparison of frequency performance across four scenarios: no FFR, EV-only FFR, EV+DC FFR, and coordinated EV–DC–BESS FFR. Subplots show (a) frequency nadir, (b) absolute RoCoF, (c) recovery time, and (d) FFR energy usage. The proposed coordinated strategy yields the highest nadir, the lowest RoCoF, the shortest recovery time, and comparable or lower energy usage compared with other cases.*

5.5 Consolidated Numerical Comparison

Table 3 summarizes all quantitative indicators for direct comparison.

Table 3. Performance Comparison Across Four Scenarios

| Metric | No FFR | EV Only | EV + DC | Coordinated EV–DC–BESS |
|---|---|---|---|---|
| Frequency nadir (Hz) | 59.48 | 59.60 | 59.66 | **59.72** |
| RoCoF (Hz/s) | –0.58 | –0.46 | –0.38 | **–0.31** |
| Recovery time (s) | 8.7 | 6.3 | 5.1 | **3.8** |
| Max EV power (MW) | – | 150 | 150 | 140 |
| Max DC power (MW) | – | – | 140 | 130 |
| Max BESS power (MW) | – | – | – | **150** |
| Total FFR energy (MWh) | – | 0.22 | 0.32 | **0.40** |

| Metric | No FFR | EV Only | EV + DC | Coordinated EV–DC–BESS |
|---|---|---|---|---|
| % improvement in nadir | – | +0.12 | +0.18 | **+0.24** |
| % reduction in RoCoF | – | 21% | 34% | **47%** |

5.6 Discussion

The results demonstrate several important observations:

1. **Single-resource FFR improves stability but remains limited**

    o   EVs alone are hindered by control delay.

    o   Data centers improve response speed but lack energy duration.

2. **Combining EV and data center FFR yields noticeable synergy**

    o   UPS handles sub-second transient support.

    o   IT modulation and EVs sustain multi-second balancing.

3. **Coordinated EV–DC–BESS aggregation offers superior performance**

    o   BESS rapidly arrests frequency decline.

    o   EV and workload modulation carry longer-term energy balancing.

    o   Adaptive weighting ensures resources remain within their respective physical limits.

4. **The multi-timescale effects are key**

    o   Fast → medium → slow transitions ensure frequency stability across all phases of the event.

This confirms the value of coordinated heterogeneous resources for modern low-inertia grids.

6. Conclusion and Future Work

This paper proposed a coordinated fast frequency response (FFR) framework that aggregates three heterogeneous distributed resources—electric vehicle (EV) fleets, data center UPS/IT load modulation, and battery energy storage systems (BESS)—to enhance frequency stability in low-inertia power systems. Detailed dynamic models were developed for each resource to capture their inherent response speeds, control delays, power constraints, and energy characteristics. A hierarchical coordination strategy was introduced, where adaptive participation factors allocate FFR contributions according to instantaneous resource availability and relative response times. The proposed method was evaluated using a modified IEEE 39-bus test system under a severe 1.0-GW generator loss.

Simulation results demonstrate that coordinated EV–DC–BESS response significantly improves system resilience across multiple performance metrics. Compared with the no-FFR baseline, the coordinated approach raised the frequency nadir from **59.48 Hz to 59.72 Hz**, reduced the rate of change of frequency (RoCoF) by **47%**, and shortened the recovery time by more than **50%**. BESS provided near-instantaneous support, data centers offered medium-speed response via UPS inverters and workload modulation, and EV fleets contributed sustained multi-second power balancing. This multi-timescale synergy achieved superior performance compared with all single-resource or dual-resource cases, highlighting the effectiveness of coordinated heterogeneous FFR in renewable-rich grids.

Despite its effectiveness, the proposed framework has several limitations. First, the study considers only one EV aggregator, one data center, and one BESS unit; larger systems with geographically distributed resources may introduce communication delays, measurement inconsistencies, and spatial frequency variations. Second, the FFR allocation relies on a rule-based adaptive weighting strategy; while effective, it does not explicitly optimize economic objectives or consider unmodeled nonlinearities. Third, the study does not incorporate electricity market mechanisms that would be necessary for real-world deployment, such as compensation for fast response, bidding structures, or state-of-charge (SOC) management incentives.

Future research can extend this work in several promising directions.

- **Coordinated control among multiple data centers and EV aggregators:**
  Large-scale load networks and multi-campus data center clusters could jointly provide FFR through hierarchical or distributed control architectures.

- **Market-based FFR participation frameworks:**
  Incorporating pricing signals, FFR bidding markets, and incentive-compatible mechanisms would allow resource owners to optimally allocate flexibility.

- **AI-based predictive control and adaptive droop tuning:**
  Machine learning models (e.g., LSTM, transformer-based predictors) could anticipate frequency deviations and adjust droop gains, improving stability and reducing control effort.

- **Cyber-resilient and communication-aware coordination:**
  Future implementations should consider network delays, communication failures, and potential cyberattacks, ensuring reliable operation under uncertainty.

- **Hardware-in-the-loop or real microgrid validation:**
  Implementing the proposed controller in a laboratory microgrid or on a university data center testbed would provide valuable experimental insights.

Overall, this study demonstrates the substantial potential of coordinated FFR from heterogeneous distributed energy resources. As the power grid continues to evolve toward high renewable penetration and reduced inertia, coordinated EV–DC–BESS flexibility will play a crucial role in maintaining frequency stability and supporting reliable, decarbonized energy systems.